\documentclass[]{spie}  

\usepackage{amsmath,amsfonts,amssymb}
\usepackage{graphicx}
\usepackage[colorlinks=true, allcolors=blue]{hyperref}

\title{An unsupervised learning-based shear wave tracking method for ultrasound elastography}

\author[a,b]{R\'emi Delaunay}
\author[a]{Yipeng Hu}
\author[a,b]{Tom Vercauteren}
\affil[a]{Wellcome/EPSRC Centre for Interventional and Surgical Sciences, University College London, Gower Street, London WC1E~6BT, UK}
\affil[b]{School of Biomedical Engineering \& Imaging Sciences, King’s College London, Strand, London WC2R~2LS, UK}

\authorinfo{Further author information: (Send correspondence to R\'emi Delaunay)\\ E-mail: remi.delaunay.17@ucl.ac.uk}

\begin{document} 
\thispagestyle{empty}
\newpage
\thispagestyle{empty} 
\vspace*{\fill} 
{\Large
This manuscript has been accepted to SPIE Medical Imaging 2022. Please use the following reference when citing the manuscript:
\vspace{0.5cm}

Remi Delaunay, Yipeng Hu, and Tom Vercauteren "An unsupervised learning-based shear wave tracking method for ultrasound elastography", Proc. SPIE 12038, Medical Imaging 2022: Ultrasonic Imaging and Tomography, 120380N; \textbf{https://doi.org/10.1117/12.2612200}
\vspace*{\fill}}

\newpage

\maketitle
\section{ABSTRACT}
Shear wave elastography involves applying a non-invasive acoustic radiation force to the tissue and imaging the induced deformation to infer its mechanical properties. This work investigates the use of convolutional neural networks to improve displacement estimation accuracy in shear wave imaging. Our training approach is completely unsupervised, which allows to learn the estimation of the induced micro-scale deformations without ground truth labels. We also present an ultrasound simulation dataset where the shear wave propagation has been simulated via finite element method. Our dataset is made publicly available along with this paper, and consists in 150 shear wave propagation simulations in both homogenous and hetegeneous media, which represents a total of 20,000 ultrasound images. We assessed the ability of our learning-based approach to characterise tissue elastic properties (i.e., Young's modulus) on our dataset and compared our results with a classical normalised cross-correlation approach.

\section{INTRODUCTION}
\label{sec:intro} 

Ultrasound elastography refers to a set of imaging techniques that characterises tissue mechanical properties after inducing tissue deformation. In shear wave elastography imaging (SWEI), a focused acoustic radiation force (ARF) is emitted by the ultrasound probe and applied into the tissue. The force induces small tissue displacements (about 10-40 $\mu m$), and generates quasi-planar shear waves that propagate through the tissue. The shear wave speed is relatively low in soft tissue (1-10 m/s), and is directly related to tissue stiffness (i.e., Young's modulus). Ultrafast plane-wave ultrasound imaging has enabled to track shear wave propagation with a high temporal resolution (up to ten thousand frames per second) and reconstruct 2D maps of tissue elasticity.

The shear wave propagation can be monitored by calculating the displacement between a reference image, acquired when the tissue is at rest, and a set of images acquired after the induced ARF. The displacement fields are usually obtained by using correlation-based approaches on the beamformed radio-frequency (RF) ultrasound data \cite{palmeri2006ultrasonic}. Another well-known method consists in calculating the shear wave propagation velocity by using a phase-shift estimator on the demodulated I/Q ultrasound data \cite{loupas1995axial}. Thereafter, the shear wave velocity is generally estimated with time-of-flight (TOF) methods, which determine the time of arrival of the shear wave at different lateral positions on the axial displacement fields \cite{rouze2012parameters}. Therefore, the accuracy of displacement measurement is crucial for shear wave estimation, which is challenging because of various sources of noise in the ultrasound data and the very small displacement magnitude.

Recent works have studied the use of neural network for displacement estimation in quasi-static elastography, where the tissue deformation is typically induced by applying an external force (i.e., palpation with a handheld transducer). Those methods are inspired by optical flow estimation techniques and consist in learning displacement estimation by minimizing a supervised loss function between the network's displacement estimates and their respective ground truth labels generated from finite element methods (FEM) \cite{tehrani2021mpwc}. Furthermore, we recently proposed an unsupervised training strategy that allows to train a network to predict a dense displacement field between a pair of ultrasound RF images. 

The advance of learning-based strategies applied to quasi-static elastography have raised a new research interest for both displacement and stiffness estimation in shear wave imaging. Chan et al., proposed a fully convolutional neural network (CNN) trained on synthetic 3-D displacement data for displacement estimation in acoustic radiation force imaging (ARFI) \cite{chan2021deep}. In addition, Ahmed et al. proposed a recurrent network to reconstruct the elasticity modulus from finite element method (FEM) tissue velocity data \cite{ahmed2021dswe}.

In this work, we investigated the use of CNNs to improve displacement estimation accuracy for tracking shear wave propagation, directly from RF ultrasound data. We considered two different network architectures proposed previously in strain elastography \cite{delaunay2021unsupervised}, and compared them with a conventional normalised cross-correlation method. The results presented were obtained from simulated ultrasound RF data. Our dataset is made publicly available along with this paper, and consists in 150 shear wave propagation simulations in both homogenous and hetegeneous media, which represents a total of 20,000 ultrasound images. To the best of our knowledge, this is the first study that explores the use of learning-based methods for ultrasonic shear wave tracking in SWEI \footnote{Open access database available on \url{https://www.synapse.org/SimForSWEI}}.

\section{Methods}
\subsection{Generating the numerical simulation dataset}

Numerical simulation was used to train our models and assess their ability to track shear wave propagation. A publicly available finite element method package was used to build our ultrasound simulation database \cite{palmeri2005finite}. Soft tissue was modeled in a mesh based on hexahedral elements, with node spacing of 0.34 $cm^3$ and size 35 x 25 x 10 mm in axial, lateral and out-of-plane directions, respectively. Each phantom presented either a uniform background or included a spherical inclusion with variable location and radius (between 1.5-5 mm). The attenuation coefficient and Poisson's ratio were kept fixed for each phantom (0.45 $Np/m$ and $0.495$, respectively), while the background's Young's modulus varied between 15 kPa and 30 kPa, with a 1.5 to 4-fold stiffer inclusion. Field-II, an ultrasound simulation software \cite{jensen1992calculation}, was used to simulate the ARF and the resulting pressure field. The ARF was focused at a depth of 19 mm and emitted for 71 $\mu s$ with a simulated linear transducer, with excitation frequency of 7 MHz and f-number of 2 (ratio between focal depth and transducer's aperture). The resulting force field was then used as input to model the resulting shear waves propagation. LS-Dyna (Ansys, Canonsburg, PA, USA), was used to generate the tissue displacement data. Perfect matching layers were positioned at the phantom's boundaries to attenuate shear wave reflections in the medium.

The ultrasound images were generated using Field-II, by simulating an ultrasound linear transducer with 7MHz central and 40 MHz sampling frequency. The speckle pattern typically observed in ultrasound imaging was obtained by randomly assigning a total of 30,000 scatterers per $cm^3$ across each digital phantom. The displaced scatterer phantoms were obtained by performing linear interpolation on the FEM displacement data. Each imaging sequence lasted 5 ms and consisted in 50 images acquired with a 10 kHz pulse repetition frequency. A total of 150 FEM simulations were performed, including 110 containing a spherical inclusion and 40 with an homogeneous background. The testing dataset is composed of 10 homogeneous and 15 inclusion phantoms. The other simulations were used for training, where three different speckle organisations were generated for each phantom, resulting in 375 different imaging sequences for training, i.e., a total of 18750 images.

\subsection{Unsupervised training strategy and network architectures}

The aim of shear wave ultrasound tracking is to find the micro-scale displacements generated by the induced acoustic radiation force. This is done by finding the set of dense displacement field (DDF) between a fixed image $I_{F}$ and a time series of moving images $I_{M(t)}$. The network is trained in an unsupervised way, by minimizing a weighted loss function composed of an image similarity and displacement regularisation term. The image similarity term corresponds to a negative local normalised cross-correlation (LNCC) function, which averages the NCC score between sliding windows sampled from the fixed image and the moving image resampled with the predicted displacement field. The regularisation term corresponds to the L1-norm of the second-order derivative of the axial predicted displacement field. The training loss can be written as follow:
\begin{equation}
Loss = LNCC(I_{F},I_{M}\circ{T}) + \alpha \sum_{i,j}\ (| \nabla^{2} u^{ax}_{i,j} | )
\end{equation}
where $T$ corresponds to the spatial transformation predicted by the network and applied to the moving image $I_{M}$ to map it in the fixed image space. The second-order derivatives of the predicted axial displacement field are noted $\nabla^{2} u^{ax}_{i,j}$.

The architecture of both neural networks used in this study has been described in details in our previous works, and are named USENet and ReUSENet for `(Recurrent) Ultrasound Elastography Network' \cite{delaunay2021unsupervised}. They both consist in an encoder-decoder neural network with skip connections. The encoder part is the same for both networks, and is composed of four down-sampling ResNet blocks. The decoder part of ReUSENet is made of upsampling convolutional long-short-term-memory (ConvLSTM) units, which take as input the encoded features as well as their temporal hidden states predicted from the previous displacement estimation. The convLSTM units are replaced by ResNet upsampling blocks for USENet. Finally, the DDF is predicted by first applying a convolution layer to the output of each upsampling blocks, and then by summing their resized outputs.

Both neural networks followed the same training procedure and were implemented in PyTorch. At each time step $t$, a pair of fixed and moving image was fed to the network to predict the corresponding DDF. Considering the large input size of each acquisition sequence (i.e., 1552x128x50 pixels), we used truncated backpropagation through time (TBTT) to update the network's weights, in order to reduce the computational load and to prevent from exploding/vanishing gradient problems. The regularisation loss weight and backpropagation step size were empirically set to $\alpha=0.02$ and $\beta=10$, respectively.

\subsection{Elasticity modulus estimation}

The shear wave speed was estimated from the displacement data by using a time-of-flight method that computes the time lag between adjacent lateral positions via cross-correlation \cite{tanter2008quantitative}. The correlation score $C$ calculated between adjacent lateral displacement profile $f_l$ and $f_{l+1}$ is calculated as follows:

\begin{equation}
    C_{l}(j) = \frac{\sum_{i=1}^{N} f_l(t_i) f_{l+1}(t_{i+j})}{[\sum_{i=1}^{N} f_l(t_i)^2 \sum_{i=1}^{N} f_l(t_{i+j})^2]^{1/2}}
\end{equation}

Where $t_i$ represents the time step of each displacement prediction. The time lag is calculated for each lateral position and corresponds to the maximum correlation function. The shear wave speed (SWS) can then be calculated as follows:

\begin{equation}
    SWS = \frac{d}{\Delta_t}
\end{equation}

Where $d$ is the distance between the adjacent lateral position and $\Delta_t$ is the time lag calculated by cross-correlation. Thereafter, the elasticity map was reconstructed by converting the shear wave velocities into Young's modulus values, and by applying a 9×9 median filter to mitigate the effects of outliers, as described in a previous study \cite{rouze2012parameters}.

\section{Experiments}

The performance of our method was evaluated on our numerical simulation testing dataset. We compared our learning-based approach with ground truth FEM displacement fields and displacement estimates obtained from a classical, normalised cross-correlation (NCC) approach \cite{palmeri2006ultrasonic}. The elasticity map was reconstructed by converting the shear wave velocities into Young's modulus values, and by applying a 9×9 median filter in order to mitigate the effects of outliers, as described in a previous study \cite{rouze2012parameters}. The performance of the compared techniques was assessed in terms of contrast-to-noise ratio (CNR) and signal-to-noise ration (SNR) of the elasticity map \cite{tehrani2021mpwc}. The Young's modulus mean absolute difference (MAE) between the prediction and ground truth estimates for the background and inclusion values were also calculated. We defined the CNR and SNR as follow:
\begin{align}
SNR = \frac{\mu}{\sigma} && CNR = \sqrt{\frac{2(\mu_b-\mu_i)^2}{\sigma_b^2 + \sigma_i^2}}
\end{align}
where $\mu$ and $\sigma$ represents respectively the mean and standard deviation of the elasticity map. Background and inclusion windows were used to compute the CNR for heterogeneous phantoms, where $\mu_b, \sigma_b$ and $\mu_i, \sigma_i$ are respectively the Young's modulus mean and standard deviation in the background and inclusion regions.

\begin{figure}[htb!]
\centering
\includegraphics[width=\textwidth]{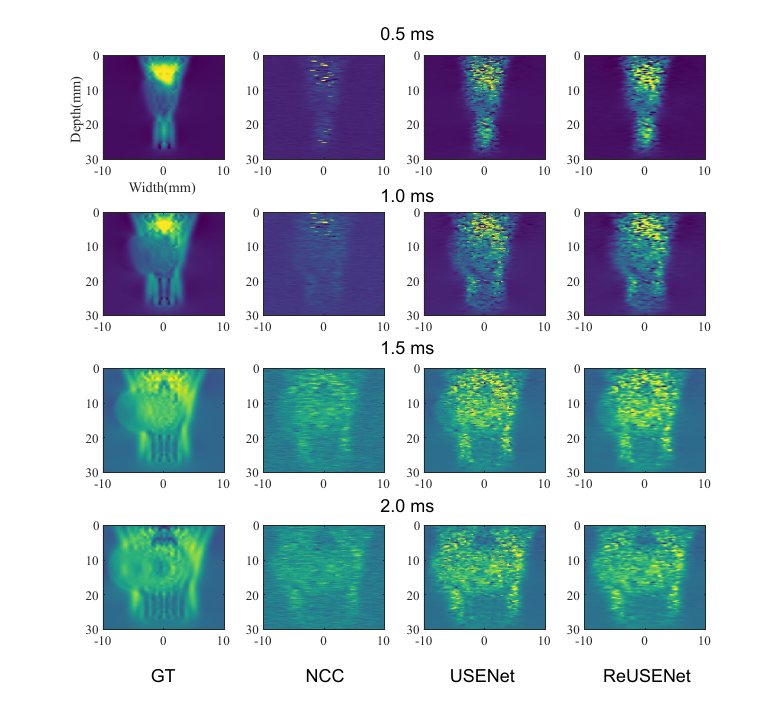}
\caption[Shear wave displacement estimated in an heterogeneous phantom]{Shear wave displacement estimated in an heterogeneous phantom. Each column corresponds to a different displacement estimation method: FEM ground truth (GT), NCC, USENet and ReUSENet. Each row corresponds to the z-score normalised displacements at different time step, from top to bottom: 0.5, 1.0, 1.5, 2.0 ms. Notice the tracked shear waves are propagating faster on the left side, indicating the presence of a stiffer inclusion.}
\label{fig:1}
\end{figure}

Figure \ref{fig:1} shows an example of z-score normalised displacements estimated from the testing dataset with ground truth displacements, NCC, USENet and ReUSENet. The shear wave propagation is shown at different time steps (0.5, 1.0, 1.5 and 2.0 ms). The left shear wave is propagating faster than the right one in the medium, which indicates that a stiffer spherical inclusion is located on the left side of the numerical phantom. 

\begin{figure}[htb!]
\centering
\includegraphics[width=\textwidth]{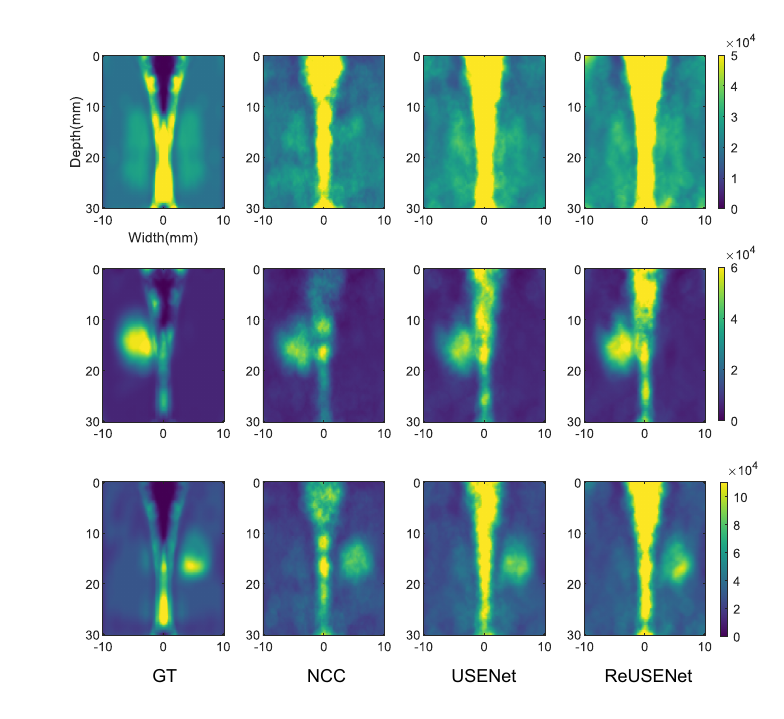}
\caption[Young's modulus elasticity maps estimated from ground truth displacement, NCC, USENet and ReUSENet]{Young's modulus elasticity maps estimated from ground truth displacement, NCC, USENet and ReUSENet. Each row display the Young's modulus map (in kPA) estimated from a different numerical phantom.}
\label{fig:2}
\end{figure}

Figure \ref{fig:2} shows three examples of elasticity map estimated from the testing dataset with ground truth displacements, NCC, USENet and ReUSENet. The first row corresponds to a numerical phantom with homogeneous background, while the two other rows show numerical phantoms that contain a stiffer inclusion. It is important to note that SWEI does not allow to characterise tissue stiffness at the excitation location \cite{tanter2008quantitative}. Therefore, the Young's modulus values displayed in the focal region are not to be taken into account for stiffness evaluation.

The CNR, SNR and Young's modulus MAE values averaged over the testing dataset for the compared methods are showed in Table \ref{tab:1}. USENet, ReUSENet and NCC methods have similar Young's modulus prediction accuracy, with slightly better results with ReUSENet for estimating the Young's modulus of the inclusion, i.e., MAE of 9.61 kPa as compared with the ground truth values. On the other hand, NCC performs slightly better in predicting the Young's modulus of the inclusion, i.e, MAE of 2.18 kPA. Our results also suggest that ReUSENet presents a better SNR and CNR (0.85, 3.90), as compared with NCC (0.82, 3.32) and USENet (0.66, 3.59).
\begin{table}[htb!]
\caption{Mean  and  standard  deviation  of  SNR, CNR and Young's modulus MAE  for the elasticity maps estimated from the simulation testing dataset (10 homogeneous background and 15 with stiffer inclusion). Results obtained from ground truth (GT), NCC, ReUSENET and USENet are showed.}
\label{tab:1}
\begin{tabular}{c|c|c|c|c}
\hline
         & \begin{tabular}[c]{@{}c@{}}SNR\\ (N=25)\end{tabular} & \begin{tabular}[c]{@{}c@{}}CNR\\ (N=15)\end{tabular} & \begin{tabular}[c]{@{}c@{}}YM target MAE (in kPa)\\ (N=15)\end{tabular} & \begin{tabular}[c]{@{}c@{}}YM background MAE (in kPa)  \\ (N = 25)\end{tabular} \\ \hline
GT       & 1.25 ($\pm 0.39$)                                         & 4.59 ($\pm 1.80$)                                       & -                                                                       & -                                                                               \\ \hline
NCC      & 0.82 ($\pm 0.21$)                                          & 3.32 ($\pm 1.91$)                                         & 12.56 ($\pm 15.22$)                                                           & \textbf{2.18 ($\pm 1.6$)}                                                            \\ \hline
USENet   & 0.66 ($\pm 0.12$)                                         & 3.59 ($\pm 1.81$)                                         & 11.22 ($\pm 13.32$)                                                          & 4.36 ($\pm 3.18$)                                                                  \\ \hline
ReUSENet & \textbf{0.85 ($\pm 0.23$)}                                 & \textbf{3.90 ($\pm 1.85$)}                               & \textbf{9.61 ($\pm 12.2$)}                                                   & 3.21 ($\pm 2.25$)                                                                    \\ \hline
\end{tabular}
\end{table}

\section{Discussion}

This study proposed the use of feed-forward and recurrent neural networks (i.e., USENet and ReUSENet) to track shear wave propagation in ultrasound elastography. We validated our method on a numerical dataset that simulated the propagation of shear waves induced by a single-push acoustic radiation force. The training methodology is completely unsupervised and ground truth displacements estimated via FEM were only used for the method comparison. We also compared our method with a classical NCC approach in terms of displacement and elasticity estimation accuracy. 

Our results suggest that learning-based methods could efficiently be used to track micro-scale deformations in SWEI. Our approach gives similar performance than the NCC method in terms of CNR, SNR and YM MAE values. Our results also indicate that ReUSENet slightly outperforms USENet. This suggests that the ability of ReUSENet to incorporate long-term dependencies with its convLSTM units improves the displacement estimation of time-series inputs. 

However, the difference in terms of displacement accuracy between ReUSENet and USENet is not as important as in our previous work on quasi-static elastography, where we showed that feed-forward networks failed to estimate large range quasi-static deformations \cite{delaunay2021unsupervised}. In shear wave imaging, the displacement generated is in the order of a few micrometres, and the ultrasound time series are acquired with a high temporal sampling rate (about 5-10 ms). Therefore, our results suggest that both feed-forward and recurrent neural networks can efficiently track shear wave displacements, and that the use of convLSTM units is better suited for ultrasound time series that exhibit larger deformations.

Finally, this work presented a publicly available ultrasound simulation dataset of 20,000 images. The ultrasound images were simulated from 150 different phantoms with either a homogenous and hetegeneous media. This dataset could be used to explore and validate further learning-based strategies for shear wave ultrasonic tracking.

\section{Acknowledgement}
This work was supported by the EPSRC [NS/A000049/1], [NS/A000050/1], [EP/L016478/1] and the Wellcome Trust [203148/Z/16/Z], [203145/Z/16/Z]. Tom Vercauteren is supported by a Medtronic/RAEng Research Chair [RCSRF1819/7/34].
\bibliography{report.bib} 
\bibliographystyle{spiebib} 

\end{document}